\documentclass[aps,
showpacs,pra,amssymb, amsmath]{revtex4}
\usepackage{amsmath,latexsym,amssymb}
\usepackage[dvips]{graphicx}
\usepackage{amsmath}
\usepackage{latexsym}
\usepackage{amssymb}
\usepackage{bm}

\begin{document}

\title{Necessary and sufficient condition for 
Greenberger-Horne-Zeilinger diagonal states
to be full $N$-partite entangled}

\author{Koji Nagata}
\affiliation{Department of Physics, Korea Advanced Institute of 
Science and Technology,
Daejeon 305-701, Korea}

\pacs{03.67.Mn, 03.65.Ca}
\date{\today}

\begin{abstract}
Recently, [{arXiv:0810.3134}] is accepted and published.
We show that any $N$-qubit state which is diagonal in the 
Greenberger-Horne-Zeilinger basis
is full $N$-qubit entangled state if and only if no partial transpose of the multiqubit state is positive with respect to any partition.
\end{abstract}

\maketitle

\section{Introduction}

Recently, \cite{NagataNakamura} is accepted and published.
Quantum information theory \cite{Nielsen,Galindo}
relies on utilizing entangled state. 
Also there is much research of nature of entangled 
states related to local realistic theories \cite{BELL,bib:Redhead,bib:Peres}.
Separable state and entangled state were defined in 1989 \cite{WERNER1}.
And it was discussed very much which state is separable or entangled \cite{DUR1,DUR2,KRAUS,KARNAS,FEI,DOHERTY,YU,CHEN}.

Peres and Horodecki provided the method of classification of a state.
A multipartite state $\rho$ has positive partial 
transposes with respect to all subsystems 
if $\rho$ is separable state \cite{PPT}.
The partial transpose of an operator on 
a Hilbert space ${H}_1\otimes{H}_2$ is defined by
\begin{eqnarray}
\left(\sum_l A^1_l\otimes A^2_l\right)^{T_1}=\sum_l{A_l^1}^{T} \otimes A_l^2,
\end{eqnarray}
where the superscript $T$ denotes transposition in the given basis.
Especially, Horodecki {\it et al}. showed \cite{PPT2} that
the condition is sufficient for quantum states to be separable on
$H= C^2\otimes C^2$ and 
$H= C^2\otimes C^3$. 
However it was shown \cite{HORODECKI} 
that the condition is not sufficient for 
quantum states to be separable in general.

For mixed high-dimensional multipartite states, the classification of 
quantum states becomes 
much more complicated.
It will be helpful to give some discussions
that determine if a given state is full multipartite entangled state.
Therefore several researches concerning to a sufficient condition of the detection of multipartite entangled state were reported (cf. \cite{TOTH}).
However, a necessary condition for a state to be full multipartite
entangled state is open question.

In this paper we provide a necessary and sufficient condition 
for a specific type of states to be full $N$-qubit entangled 
state.
Here, we shall investigate the property of any $N$-qubit state which is 
diagonal in the Greenberger-Horne-Zeilinger (GHZ) \cite{GHZ,GHZ2} basis.
It turns out that such a GHZ diagonal state 
is full $N$-qubit entangled state if and only if no partial transpose of the multiqubit state is positive with respect to any partition.

Our result is generalization of the result presented 
in Refs.~\cite{DUR1,DUR2} about a subset of the family of GHZ
diagonal states.
Our result is for the general class of GHZ diagonal states.

First we note a definition of full $N$-qubit entangled state as follows:

Consider a partition of $N$-qubit
system ${\bf N}_N=\{1,2,\ldots,N\}$ into $2$
nonempty and disjoint subsets 
$\alpha_1$ and $\alpha_2$, where 
$|\alpha_1|+|\alpha_2|=N$, 
to which we refer as a bi-partite split of the system.
Let us consider a density operator $W$ on 
$H=\otimes_{j=1}^N H_j$, where $H_{j}$ represents
the subspace with respect to particle $j$.

A density operator $W$ may be called bi-separable with respect to a 
partition $\alpha_1,\alpha_2$ iff
it can be written as
\begin{eqnarray}
W=\sum_{l} p_l 
\left(\otimes^2_{i=1}W_l^{\alpha_i}\right), \left(p_l\geq 0,\sum_l p_l=1\right)
\label{ksep},
\end{eqnarray}
where $W^{\alpha_i}_l,\forall l$ are the density operators 
on the subspace 
$\otimes_{j\in\alpha_i}{H}_{j}$.
If a density matrix can be written by a density matrix of 
the form (\ref{ksep}), it is bi-separable with respect to a 
partition $\alpha_1,\alpha_2$.
If the density matrix cannot be written by a density matrix of the form
(\ref{ksep}) with respect to any partition, 
it is full $N$-qubit entangled state.


\section{Necessary and sufficient condition for Bell diagonal states
to be entangled}

In this section, we review 
the necessary and sufficient condition for Bell diagonal states
to be entangled as follows.
Assume Bell diagonal states as
\begin{eqnarray}
\rho=\delta_1|\psi^+\rangle\langle\psi^+|
+\delta_2|\psi^-\rangle\langle\psi^-|
+\delta_3|\phi^+\rangle\langle\phi^+|
+\delta_4|\phi^-\rangle\langle\phi^-|\label{Bell}
\end{eqnarray}
where $A=\delta_1+\delta_2+\delta_3+\delta_4=1$ and
\begin{eqnarray}
&&|\psi^+\rangle=\frac{1}{\sqrt{2}}(|1,1\rangle+|0,0\rangle),\nonumber\\
&&|\psi^-\rangle=\frac{1}{\sqrt{2}}(|1,1\rangle-|0,0\rangle),\nonumber\\
&&|\phi^+\rangle=\frac{1}{\sqrt{2}}(|1,0\rangle+|0,1\rangle),\nonumber\\
&&|\phi^-\rangle=\frac{1}{\sqrt{2}}(|1,0\rangle-|0,1\rangle).\label{set}
\end{eqnarray}
We know that $\rho^{T}\geq 0$ is the necessary and sufficient condition
for the density matrix $\rho$ to be separable since
positivity of partial transpose is the necessary and sufficient condition 
for quantum states to be separable on $C^2\otimes C^2$ \cite{PPT2}.
The eigenvalues of partial transpose $\rho^T$ are 
\begin{eqnarray}
&&B=\delta_1-\delta_2+\delta_3+\delta_4,\nonumber\\
&&C=\delta_1+\delta_2-\delta_3+\delta_4,\nonumber\\
&&D=\delta_1+\delta_2+\delta_3-\delta_4,\nonumber\\
&&E=-\delta_1+\delta_2+\delta_3+\delta_4.
\end{eqnarray}
Thus, if and only if
\begin{eqnarray}
\rho^{T}\geq 0\Leftrightarrow
B,C,D,E\geq 0,
\end{eqnarray}
$\rho$ is separable state.

\section{Necessary and sufficient condition for GHZ diagonal states
to be full $N$-partite entangled}

Consider a subset $\alpha\subset {\bf N}_N$ and a
density operator $W$ acting on ${H}$,
let $W^{T_{\alpha}}$ 
denote the partial transpose of all sites belonging to $\alpha$.
Let ${P}$ denote a family of sets, which consists of all unions of 
$\alpha_1,\alpha_2$ together with the empty set, so that
${P}$ has $2^2$ elements.
A density operator $W$ 
may be called $2$-positive partial transpose ($2$-PPT) 
with respect to this specific partition iff
$W^{T_{\alpha}}\geq 0$ for all 
$\alpha\in{P}$.

Total transposition preserves the spectrum of a bipartite state. Moreover, total transposition with respect to one side of the partition followed by total transposition corresponds to partial transposition on the other side. Thus, in the partial transposition test, it is sufficient to consider transposition with respect to only the sites belonging to $\alpha$, (moreover, this means that checking the positivity of the partial transpose with respect to $\alpha=\emptyset$ or $\alpha={\bf N}_N$ would correspond just to check the positivity of the density matrix, which is positive by definition).

Clearly if a density operator $W$ is not $2$-PPT 
with respect to any partition, the state $W$
should be full $N$-qubit entangled state.

Let $\beta$ be a subset $\beta\subset{\bf N}_N$
and $l(\beta)$ be an integer $l_1\cdots l_N$ in binary notation
with $l_m=1$ for $m\in \beta$ and $l_m=0$ otherwise 
$(0\le l(\beta)\le 2^{N}-1)$ \cite{note}.

Using an integer $l(\beta)$ and a subset $\alpha\subset{\bf N}_N\wedge\alpha\neq\emptyset$, 
we introduce two vectors as 
$\{\otimes_{m\in \alpha}|l_m\rangle,
\otimes_{m\in \alpha}|l_m\oplus 1\rangle\}=
\{|B_{\beta}(\alpha)\rangle,|\overline{B_{\beta}(\alpha)}\rangle\}$.
Here $l_m\oplus 1$ is 
the bitwise XOR (exclusive OR) of $l_m$ and $1$.
Hence, $l_m\oplus 1=0$ if $l_m=1$. And $l_m\oplus 1=1$ if $l_m=0$.
Here both vectors $|B_{\beta}(\alpha)\rangle$ 
and $|\overline{B_{\beta}(\alpha)}\rangle$ 
are acting on the subspace $\otimes_{m\in \alpha}H_m$.
And $\langle{B_{\beta}(\alpha)}|\overline{B_{\beta}(\alpha)}\rangle=0$
holds for every subset $\alpha,\beta$.
For a given $\alpha, \beta$, two vectors 
($|B_{\beta}(\alpha)\rangle,|\overline{B_{\beta}(\alpha)}\rangle$)
form a two-dimentional space.

Again, consider a partition of $N$-qubit
system ${\bf N}_N=\{1,2,\ldots,N\}$ into $2$
nonempty and disjoint subsets 
$\alpha_1$ and $\alpha_2$, where 
$|\alpha_1|+|\alpha_2|=N$, 
to which we referred as a bi-partite split of the system.

The orthonormal GHZ basis for $2^N$-dimentional space is covered by the following family of 
states given by
\begin{eqnarray}
|\Psi_{\beta}^{\pm}\rangle=
\frac{1}{\sqrt{2}}(|B_{\beta}(\alpha_1)|{B_{\beta}(\alpha_2)}\rangle\pm 
|\overline{B_{\beta}(\alpha_1)}\rangle|\overline{B_{\beta}(\alpha_2)}\rangle),\beta\subset{\bf N}_N.
\end{eqnarray}
Please notice that 
\begin{eqnarray}
|\Psi_{\beta}^{\pm}\rangle\langle \Psi_{\beta}^{\pm}|
=|\Psi_{{\bf N}_N\backslash \beta}^{\pm}\rangle
\langle |\Psi_{{\bf N}_N\backslash \beta}^{\pm}|.
\end{eqnarray}
Therefore, to make the GHZ basis for $2^N$-dimentional space, it is sufficient to consider
$2^{N-1}$ kinds of subset ($\beta$) even though 
the cardinal number of subsets of ${\bf N}_N$ is $2^N$.

In what follows, we shall show that {\it any $N$-qubit state which is 
diagonal in the GHZ basis is full $N$-qubit entangled state if and only if no partial transpose of the state is positive with respect to any partition.}

Let us consider such a multiqubit density operator $X$ as
\begin{eqnarray}
X=\frac{1}{2}\sum_{\beta\subset{\bf N}_N}
(\lambda^+_{\beta}|\Psi_{\beta}^{+}\rangle
\langle\Psi_{\beta}^{+}|
+\lambda^-_{\beta}|\Psi_{\beta}^{-}\rangle
\langle \Psi_{\beta}^{-}|).
\end{eqnarray}
Of course 
$(1/2)\sum_{\beta\subset{\bf N}_N}
(\lambda^+_{\beta}+\lambda^-_{\beta})=1,~\lambda^+_{\beta}\geq 0,~\lambda^-_{\beta}\geq 0$.
The values of the positive coefficients of $X$, i.e., all
$\lambda$ are 
\begin{eqnarray}
\lambda_{\beta}^{\pm}=\langle \Psi^{\pm}_{\beta}|X|\Psi^{\pm}_{\beta}\rangle.
\end{eqnarray}

In what follows, we shall show that 
{\it $X$ is bi-separable with respect to a partition $\alpha_1$ 
and $\alpha_2$ if and only if the partial transpose with respect to 
the partition is positive, $X^{T_{\alpha_1}}\geq 0$}.

We introduce a positive operator as
\begin{eqnarray}
&&Y_{\beta,\alpha_1,\alpha_2}=
\lambda^+_{\beta}|\Psi_{\beta}^{+}\rangle
\langle\Psi_{\beta}^{+}|
+\lambda^-_{\beta}|\Psi_{\beta}^{-}\rangle
\langle \Psi_{\beta}^{-}|
+\eta^+_{\beta}|\Phi_{\beta}^{+}\rangle
\langle\Phi_{\beta}^{+}|
+\eta^-_{\beta}|\Phi_{\beta}^{-}\rangle
\langle \Phi_{\beta}^{-}|\label{quasi}
\end{eqnarray}
where $|\Phi_{\beta}^{\pm}\rangle$ is given by
\begin{eqnarray}
|\Phi_{\beta}^{\pm}\rangle=
\frac{1}{\sqrt{2}}(|B_{\beta}(\alpha_1)\rangle
|\overline{B_{\beta}(\alpha_2)}\rangle\pm 
|\overline{B_{\beta}(\alpha_1)}\rangle|{B_{\beta}(\alpha_2)}\rangle).
\end{eqnarray}
Here, 
\begin{eqnarray}
\eta_{\beta}^{\pm}
=\langle \Phi^{\pm}_{\beta}|X|\Phi^{\pm}_{\beta}\rangle.
\end{eqnarray}
The operator $Y_{\beta,\alpha_1,\alpha_2}$ is acting on subspace spanned by 
\begin{eqnarray}
|\Psi_{\beta}^{+}\rangle=
\frac{1}{\sqrt{2}}(|B_{\beta}(\alpha_1)\rangle
|{B_{\beta}(\alpha_2)}\rangle+ 
|\overline{B_{\beta}(\alpha_1)}\rangle|\overline{B_{\beta}(\alpha_2)}\rangle),\nonumber\\
|\Psi_{\beta}^{-}\rangle=
\frac{1}{\sqrt{2}}(|B_{\beta}(\alpha_1)\rangle|{B_{\beta}(\alpha_2)}\rangle-
|\overline{B_{\beta}(\alpha_1)}\rangle|\overline{B_{\beta}(\alpha_2)}\rangle),\nonumber\\
|\Phi_{\beta}^{+}\rangle=
\frac{1}{\sqrt{2}}(|B_{\beta}(\alpha_1)\rangle|\overline{B_{\beta}(\alpha_2)}\rangle+ 
|\overline{B_{\beta}(\alpha_1)}\rangle
|{B_{\beta}(\alpha_2)}\rangle),\nonumber\\
|\Phi_{\beta}^{-}\rangle=
\frac{1}{\sqrt{2}}(|B_{\beta}(\alpha_1)\rangle|\overline{B_{\beta}(\alpha_2)}\rangle- 
|\overline{B_{\beta}(\alpha_1)}\rangle|{B_{\beta}(\alpha_2)}\rangle).
\end{eqnarray}
Clearly, the above set of four vectors is 
analogous to the set of four vectors presented in (\ref{set}).
Hence, we see that a generic bipartite state is mapped in 
an operator as in
Eq.~(\ref{quasi}).
We thus see that an operator in
Eq.~(\ref{quasi}) described in $N$-qubit system 
is analogous to a quantum state 
in Eq.~(\ref{Bell}) described in
two-qubit system.
But normalization factor may be different.
That is, $A_{\beta}=\lambda^+_{\beta}+
\lambda^-_{\beta}
+\eta^+_{\beta}
+\eta^-_{\beta}\leq 1$.
In order to follow the analogy to two-qubit Bell diagonal states, 
we introduce the following notations as
\begin{eqnarray}
&&B_{\beta}=
\lambda^+_{\beta}
-\lambda^-_{\beta}
+\eta^+_{\beta}
+\eta^-_{\beta},\nonumber\\
&&C_{\beta}=\lambda^+_{\beta}+
\lambda^-_{\beta}
-\eta^+_{\beta}
+\eta^-_{\beta},\nonumber\\
&&D_{\beta}
=\lambda^+_{\beta}
+\lambda^-_{\beta}
+\eta^+_{\beta}
-\eta^-_{\beta},\nonumber\\
&&E_{\beta}
=-\lambda^+_{\beta}+
\lambda^-_{\beta}
+\eta^+_{\beta}
+\eta^-_{\beta}.
\end{eqnarray}
Then, if and only if
\begin{eqnarray}
Y_{\beta,\alpha_1,\alpha_2}^{T_{\alpha_1}}\geq 0\Leftrightarrow
B_{\beta},
C_{\beta},
D_{\beta},
E_{\beta}\geq 0\label{NS}
\end{eqnarray}
$Y_{\beta,\alpha_1,\alpha_2}/A_{\beta}$ is bi-separable 
with respect to a partition $\alpha_1$ 
and $\alpha_2$ since
positivity of partial transpose is the necessary and sufficient condition 
for quantum states to be separable on $C^2\otimes C^2$ \cite{PPT2}.
That is, we have 
considered the positive operator $Y_{\beta,\alpha_1,\alpha_2}$ 
as a two-qubit state.
Hence, if and only if the condition (\ref{NS}) holds, $Y_{\beta,\alpha_1,\alpha_2}$ 
can be written as
\begin{eqnarray}
Y_{\beta,\alpha_1,\alpha_2}=\sum_{l} q_l 
\left(W_l^{\alpha_1}\otimes W_l^{\alpha_2}\right), \left(q_l\geq 0,\sum_l q_l=A_{\beta}\leq 1\right).
\end{eqnarray}
One has
\begin{eqnarray}
X=(1/4)\bigotimes_{\beta\subset{\bf N}_N}
Y_{\beta,\alpha_1,\alpha_2}.
\end{eqnarray}
Hence, we have
\begin{eqnarray}
X^{T_{\alpha_1}}=
(1/4)\bigotimes_{\beta\subset{\bf N}_N}
Y_{\beta,\alpha_1,\alpha_2}^{T_{\alpha_1}}.
\end{eqnarray}
This implies $X^{T_{\alpha_1}}\geq 0\Leftrightarrow 
Y_{\beta,\alpha_1,\alpha_2}^{T_{\alpha_1}}\geq 0:\forall 
\beta\subset{\bf N}_N$.
Hence, $X$ is bi-separable with respect to a partition $\alpha_1$ 
and $\alpha_2$ if and only if the partial transpose with respect to 
the partition is positive, $X^{T_{\alpha_1}}\geq 0$.

On the other hand, $X$ is full $N$-qubit entangled state if there is no 
partition $\alpha_1,\alpha_2$ such that $X^{T_{\alpha_1}}\geq 0$.

\section{Summary}

In summary, we have investigated a specific type of $N$-qubit states and 
showed that any $N$-qubit state which is 
diagonal in the GHZ basis
is bi-separable for a partition if and only if 
partial transpose of the multiqubit state 
is positive with respect to the partition.
This implies that there is no positive partial transpose for any subsystems if and only if the GHZ diagonal $N$-qubit state is full $N$-qubit entangled state.

We cannot generalize our arguments into general multiqubit states, 
just because there
are PPT entangled states of $4$ qubits. For instance,
in Ref.~\cite{bound}, a $4$-qubit state
is discussed which is positive under transposition on qubits
$3,4$, but entangled with respect to the two-component partition
$\{1,2,3,4\}=\{1,2\}\cup\{3,4\}$.

\section*{Acknowledgments}
The authors thank G. T\'oth and O. G\"uhne for valuable comments.
This work has been
supported by Frontier Basic Research Programs at KAIST and K.N. is
supported by a BK21 research grant.


\begin{thebibliography}{00}
\bibitem{note}
Therefore using the following formula
$
\sum_{i=0}^{N-1} 2^{i}=2^N-1,
$
we can see that
$
l(\beta)=\sum_{i=0}^{N-1} l_{N-i}(\beta)2^{i}.
$
\end{thebibliography}
\end{document}